\documentclass[floats,floatfix,showpacs,amssymb,prd,superscriptaddress,twocolumn,aps,nofootinbib]{revtex4-1}
\usepackage{graphicx, epsfig, amssymb} 
\usepackage{amsmath, amsfonts}
\usepackage{bm} 

\usepackage[linktocpage]{hyperref}
\usepackage[caption=false]{subfig}
\usepackage[usenames]{color}

\DeclareMathOperator\erf{erf}

\newcommand{\dif}{{\rm d}}

\def\beq{\begin{equation}}
\def\eeq{\end{equation}}
\def\beqa{\begin{eqnarray}}
\def\eeqa{\end{eqnarray}}

\def\non{\nonumber}

\begin{document}
\title{\large Collapse of self-interacting fields 
in asymptotically flat spacetimes:\\do self-interactions render Minkowski spacetime unstable?}
\author{Hirotada Okawa}
\affiliation{CENTRA, Departamento de F\'{\i}sica, Instituto Superior T\'ecnico, Universidade de Lisboa,
Avenida Rovisco Pais 1, 1049 Lisboa, Portugal}

\author{Vitor Cardoso} 
\affiliation{CENTRA, Departamento de F\'{\i}sica, Instituto Superior T\'ecnico, Universidade de Lisboa,
Avenida Rovisco Pais 1, 1049 Lisboa, Portugal}
\affiliation{Perimeter Institute for Theoretical Physics Waterloo, Ontario N2J 2W9, Canada}

\author{Paolo Pani} 
\affiliation{CENTRA, Departamento de F\'{\i}sica, Instituto Superior T\'ecnico, Universidade de Lisboa,
Avenida Rovisco Pais 1, 1049 Lisboa, Portugal}
\affiliation{Institute for Theory and Computation, Harvard-Smithsonian
CfA, 60 Garden Street, Cambridge, MA 02138, USA}

\date{\today} 

\begin{abstract} 
The nonlinear instability of anti-de Sitter spacetime has recently been established with the striking result that generic initial data collapses to form black holes.
This outcome suggests that confined matter might generically collapse, and that collapse could only be halted -- at most -- by nonlinear bound states.
Here we provide evidence that such mechanism can operate even in asymptotically flat spacetimes, by studying the evolution of the Einstein-Klein-Gordon system for a self-interacting scalar field. 
We show that (i) configurations which do not collapse promptly can do so after successive reflections off the potential barrier, but (ii) that at intermediate amplitudes and Compton wavelengths, collapse to black holes is
replaced by the appearance of oscillating soliton stars, or ``oscillatons''. 
Finally, (iii) for very small initial amplitudes, the field disperses away in a manner consistent with power-law tails of massive fields. Minkowski is stable against gravitational collapse. 
Our results provide one further piece to the rich phenomenology of gravitational collapse and show the important interplay between bound states, blueshift, dissipation and confinement effects.
\end{abstract}

\maketitle
\section{Introduction}
The nonlinear stability of spacetimes against gravitational collapse is a highly nontrivial and important problem.
Gravity is attractive and tends to clump things together, but the global structure of spacetime may play an important role, for example by allowing energy to be dispersed away. The nonlinear stability of Minkowski spacetime was established rigorously decades ago~\cite{christodouloubook}. 
In this case any arbitrarily small initial perturbation eventually disperses to infinity. As the amplitude of the initial data is tuned up,
collapse eventually ensues, driven by nonlinear gravitational effects~\cite{Choptuik:1992jv}.

Somewhat surprisingly, it was recently shown through convincing numerical results that anti-de Sitter (AdS) spacetime is unstable against gravitational collapse:
any initial disturbance, however small, eventually forms black holes~\cite{Bizon:2011gg}. The physics of the process is not yet fully understood, but it hinges on two key ingredients: confinement of matter by the AdS boundary and the attractive character of gravity. Because in AdS spacetime the perturbation cannot simply disperse away, it is forced to continuously interact nonlinearly, eventually collapsing. 
At perturbative level, this effect seems to hinge on a weakly turbulent instability that focuses the energy as the system evolves~\cite{Bizon:2011gg,Dias:2011ss}. 
It was also found that apparent exceptions to this rule exist if the initial data is tuned to form possible nonlinear bound states~\cite{Dias:2011ss,Buchel:2013uba,Maliborski:2013jca,Maliborski:2013ula,Dias:2012tq}, but these results all leave open the possibility of collapse on very large time scales not covered by the simulations and where nonlinearities might play a dominant role.

These ingredients for collapse can, in one guise or another, be also active in asymptotically flat spacetime. For example gravitational potential wells, such as those experienced within	 stars or in the early universe, can confine the initial perturbations and force them to interact nonlinearly. Another example concerns the evolution
of free massive fundamental fields, where the mass term is known to provide low-frequency confinement and allow for the existence of bound states such as soliton stars~\cite{Seidel:1991zh,Seidel:1993zk,Page:2003rd} or long-lived condensates around black holes~\cite{Cardoso:2005vk,Dolan:2007mj,Dolan:2012yt,Pani:2012vp,Pani:2012bp,Witek:2012tr}. Because confinement is a key player in the nonlinear, turbulent instability of ``boxed spacetimes'' \cite{Bizon:2011gg,Maliborski:2013ula}, we are left with the exciting and troublesome possibility that gravitational collapse is a generic rule --~rather than an exception~-- in our Universe. Motivated by this, we revisit an old problem on the gravitational collapse of self-interacting scalar fields~\cite{Goncalves:1997qp,Brady:1997fj}.

\section{Setup}
We consider the Einstein-Klein-Gordon theory for a massive scalar field $\Phi$ (we work in $G=c=\hbar=1$ units) 
\begin{align}
\label{eq:action}
S = & \int d^4x \sqrt{-g} 
      \left( \frac{R}{16\pi} 
            -\frac{1}{2}\nabla^{\mu}\Phi\nabla_{\mu}\Phi -\frac{1}{2}\mu^2\Phi^2 \right)\,,
\end{align}
although our main results and conclusions seem to extend also to self-interacting fields with $\Phi^4$ interactions.
We focus on spherically symmetric spacetimes. In the Arnowitt-Deser-Misner (ADM) decomposition, the geometry is described by
\begin{subequations}
 \label{eq:Ansatz_SS}
  \begin{align}
   \label{eq:Ansatz_SS_met}
   \dif s^2 =& -\alpha^2\dif t^2 +\psi^4\eta_{ij}\dif x^i\dif x^j\,,\\
   \label{eq:Ansatz_SS_ext}
   K_{ij} =& \frac{1}{3}\psi^4\eta_{ij}K\,,
  \end{align}
\end{subequations}
where $\eta_{ij}$ is the Minkowski 3-metric in spherical coordinates and $K_{ij}$ is the extrinsic curvature of the conformally flat metric $\gamma_{ij}=\psi^4\eta_{ij}$.

The equations of motion yield the constraints
  \begin{align}
   &\frac{2}{3}K^2-\frac{(8r^2\psi_{,r})_{,r}}{r^2\psi^5}-8\pi \left[\Pi^2+\psi^{-4}\Phi_{,r}^2+\mu^2\Phi^2\right] =\; 0\,,\nonumber\\
   \label{eq:Constr_SS}
   &\frac{2}{3}K_{,r} +8\pi\Pi\Phi_{,r} =\; 0\,,
  \end{align}
and the evolution equations
  \begin{align}
   \partial_t\psi_0 =& -\frac{1}{6}\alpha\left(1+\psi_0\right) K,\quad
   \partial_t\psi_2 = -\frac{1}{6}\alpha\psi_2 K\,,\nonumber\\
   \partial_t K =&-\psi^{-4}\alpha_{,rr} -2\psi^{-5}\psi_{,r}\alpha_{,r}
   -\frac{2\alpha_{,r}}{r\psi^4} +\frac{1}{3}\alpha K^2\nonumber\\
   &+4\pi\alpha\left(2\Pi^2 -\mu^2\Phi^2\right)\,,\nonumber\\
   \partial_t\Phi =& -\alpha\Pi\,,\nonumber\\
   \partial_t\Pi =& \alpha\Pi K
   -\psi^{-4}\alpha_{,r}\Phi_{,r} -\alpha\psi^{-4}\Phi_{,rr}
   -2\alpha\psi^{-5}\psi_{,r}\Phi_{,r}\nonumber\\
   \label{eq:evol_SS}
   & -\frac{2\alpha\Phi_{,r}}{r\psi^4}+\alpha\mu^2\Phi\,,
  \end{align}
where $\Pi$ is the momentum conjugate of $\Phi$ and we decomposed the conformal factor as $\psi \equiv\; 1 +\psi_0 +\psi_2/r^2$.

Regularity at the origin of the coordinates is achieved in a way similar to that reported in Refs.~\cite{Bizon:2011gg,Maliborski:2013via}, i.e. by noting that $1/r$ terms in our evolution equations can be
naturally regularized by $\psi^4$. 
Gauge equation is implemented similarly to Refs.~\cite{Bona:1994dr,Arbona:1998hu,Alcubierre:2004gn}.
Unlike many previous studies on gravitational collapse, the ansatz for the metric allows us to follow 
the collapse through horizon formation.

We implemented the above equations in a One Dimensional Numerical Relativity code (ODIN), where
discretization of spatial derivatives is obtained with 4th order accurate stencils.
Integration in time is done with a 4th order accurate Runge-Kutta method. Parallelization is realized with OpenMP (see Ref.~\cite{Okawa:2013afa} for details).

The outer numerical boundary is placed sufficiently far away as to be causally disconnected from the region under study, and 
each of the simulations we discuss was stopped before spurious reflections from the outer boundary can contaminate the results. 
We have verified {\it a posteriori} that the grid size is also larger than any characteristic wavelength showing up in our results.
Our results are convergent and stable when the grid size is varied.

\subsection{Initial data}
The construction of constraint-satisfying initial data is a delicate issue. Here we provide {\it analytic}
initial data that solves the constraints~\eqref{eq:Constr_SS}. The method can be generalized to less symmetric configurations
and more general physical systems. The details are presented in Ref.~\cite{Okawa:bombs_2013}.
We choose $\Phi=0$, the maximal slicing condition $K=0$, and we take the ansatz
\begin{align}
 \Pi =& \frac{A}{2\pi}\psi^{-\frac{5}{2}}\exp\left\{-(r-r_0)^2/w^2\right\},
 \quad
 \psi = 1 +\frac{u(r)}{\sqrt{4\pi}r}\,,\nonumber
\end{align}
where $A,\, r_0$ and $w$ are constants denoting the amplitude, location and width of the initial scalar pulse.
With this ansatz, the Hamiltonian constraint can be reduced to the ordinary differential equation
\begin{eqnarray}
u''(r) + \frac{A^2r}{\sqrt{4\pi}}\exp\left\{-(r-r_0)^2/w^2\right\}=0\,.\nonumber
\end{eqnarray}
A particular solution, which is regular at infinity, is
\begin{eqnarray}
\label{eq:u0}
 u_{0}(r)&= & A^2w\frac{w^2-4r_0(r-r_0)}{16\sqrt{2}}\left[\erf{\left({\sqrt{2}(r-r_0)}/{w}\right)}-1\right]\non\\
 &-&A^2\frac{r_0w^2}{8\sqrt{\pi}}\exp\left\{-2(r-r_0)^2/w^2\right\}+{\rm const}\,.
\end{eqnarray}
We are of course free to add any arbitrary constant to this solution, and we will do so
to guarantee regularity at the origin and a finite initial ADM energy, $M_0(A,r_0,w)=-u_0(0)/\sqrt{\pi}$.
Thus, we take as initial data the expressions
%
\begin{align}
 \psi_0 =& -\frac{A^2r_0w}{8\sqrt{2\pi}}\left[\erf{\left({\sqrt{2}(r-r_0)}/{w}\right)}-1\right]\,,\nonumber\\
 \frac{\psi_2}{r^2} =& \frac{u_{0}(r)-u_{0}(0)}{\sqrt{4\pi}r}
  +\frac{A^2 r_0w}{8\sqrt{2\pi}}\left[\erf{\left({\sqrt{2}(r-r_0)}/{w}\right)}-1\right]\,,\nonumber\\
 \Pi =&\frac{A}{2\pi}\psi^{-\frac{5}{2}}\exp\left\{-(r-r_0)^2/w^2\right\}\,, \nonumber\\
 \label{eq:ID_MinkowskiBG}
 \Phi =&0\,,\quad K =0,\quad \alpha = 1\,.
\end{align}
\begin{figure}[ht]
 \psfig{file=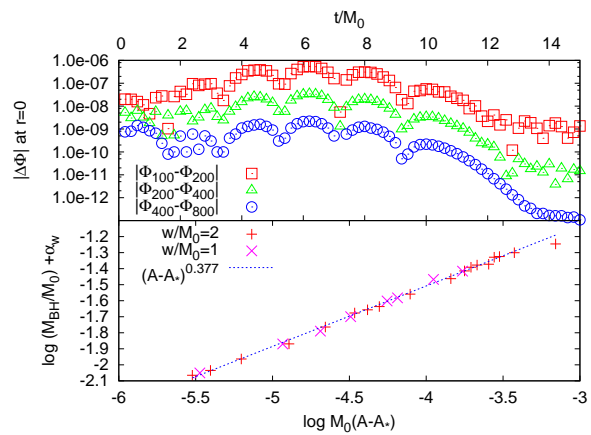,width=8cm}
 \caption[]{Top panel: convergence test of ODIN in the massless case, using as parameters
 $AM_0=0.01,\, r_0=5M_0,\,w=2M_0$ and grid size $20M_0$. 
 We monitor the scalar field $\Phi$ at the origin for different resolutions, $dr/M_0=20/100,20/200,20/400$ and $20/800$.
 As expected for 4th-order convergence, the errors decrease by a factor $\sim16$.
 Bottom panel: critical behavior of a massless field with different initial pulse widths. 
 For this choice of parameters, the critical amplitude $A_{*}$ reads
 $0.3955,0.4058$ for $w/M_0=2,1$, respectively.
 In the vertical axis, $\alpha_w$ represents the corresponding offset parameter.
 }
 \label{fig:codetest}
\end{figure}
%

\section{Type-I and type-II collapse}
%
\begin{figure}[ht]
\begin{tabular}{c}
\psfig{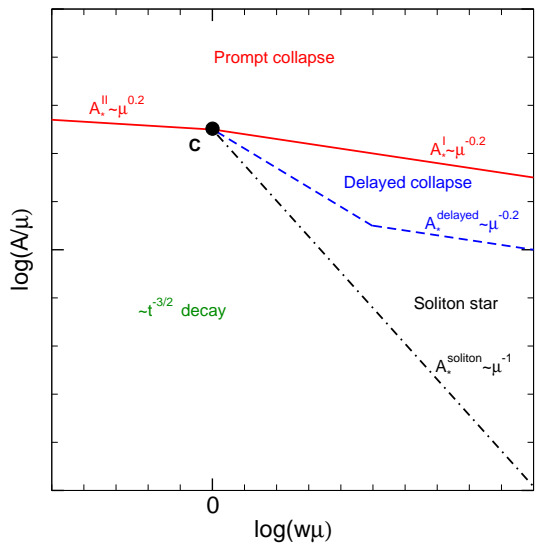}
\end{tabular}
\caption[]{Qualitative phase diagram for the spherically symmetric collapse of a massive scalar field in the $(A_0/\mu,w\mu)$ plane (in log scale). The triple point separating black-hole formation, soliton stars and power-law decay is marked by a black circle. Threshold lines refer to Eqs.~\eqref{promptII}--\eqref{eq:threshold_soliton} with $w$ set to unity for clarity.
}
\label{fig:diagram}
\end{figure}
The collapse of a massive scalar field was found to fall in two possible regimes,
depending on the width-to-Compton wavelength ratio $w\mu$~\cite{Goncalves:1997qp,Brady:1997fj}.
For small $w\mu$, the collapse generically proceeds in a qualitatively similar way to massless fields,
displaying what was termed type-II behavior: the subsequent evolution of large amplitude initial data
eventually gives rise to black-hole formation; as the initial amplitude is decreased, the black-hole size decreases and 
criticality~\cite{Choptuik:1992jv} is approached at vanishingly small black-hole mass and finite amplitude for the scalar field. Close to criticality, the black-hole mass satisfies $M_{\rm BH}/M_0=C (A-A_*)^{\gamma}$. 
Type-I collapse occurs when $w\mu\gg1$, with the black-hole mass function near criticality being discontinuous~\cite{Brady:1997fj}.

We have performed a convergence test, summarized in Fig.~\ref{fig:codetest}, 
reproducing the expected 4th-order convergence along with a width $w-$ and $r_0-$independent critical exponent $\gamma\sim 0.377$ in the massless case,
in agreement to that reported in the original work~\cite{Choptuik:1992jv}.
However, for $\mu\neq0$, we have found novel features and a much richer phenomenology. There exist several distinct phase transitions, which are governed by the initial conditions and by the mass term $\mu$.
We have performed an extensive search of over 500 simulations (whose typical resolution is $dr/M_0=1/500$) in the
entire parameter space and computed the threshold lines that separate
each phase. The corresponding $(A_0/\mu,w\mu)$ phase diagram is depicted in Fig.~\ref{fig:diagram},
which to the best of our knowledge summarizes for the first time the possible outcome of the evolution of self-interacting scalar fields.
This phase diagram also is consistent with previous studies on the subject~\cite{Brady:1997fj}.
\section{Phases of massive scalar field collapse}
The various phases can be understood by fixing the coupling $w\mu$ and slowly decreasing the amplitude, i.e. by moving on a vertical line starting from the uppermost part of Fig.~\ref{fig:diagram}.

The type-II region shows a single phase transition, namely from black-hole formation at large amplitude to power-law decay at small amplitude. The interface between these two phases is defined by the separatrix
\begin{equation}
 A>A_{*}^{\rm II}\sim \mu (w\mu)^{-0.8} \,. \label{promptII}
\end{equation}
The critical amplitude depends very mildly on $\mu$ and connects smoothly to a constant value in the $\mu\to 0$ limit. The behavior near the threshold is similar to Choptuik's critical collapse~\cite{Choptuik:1992jv}.

On the other hand the gravitational collapse of fields which extend
sufficiently far beyond their Compton wavelength is drastically different.
If the mass term is sufficiently large, our results indicate that prompt collapse occurs for initial data satisfying
\begin{equation}
 A>A_{*}^{\rm I}\sim \mu (w\mu)^{-1.2} \qquad w\mu\gg1\,. \label{prompt}
\end{equation}
A WKB analysis of this regime shows that the scalar collapse is similar to pressureless dust and agrees qualitatively with our findings~\cite{Goncalves:1997qp}.

\begin{figure}[ht]
\psfig{file=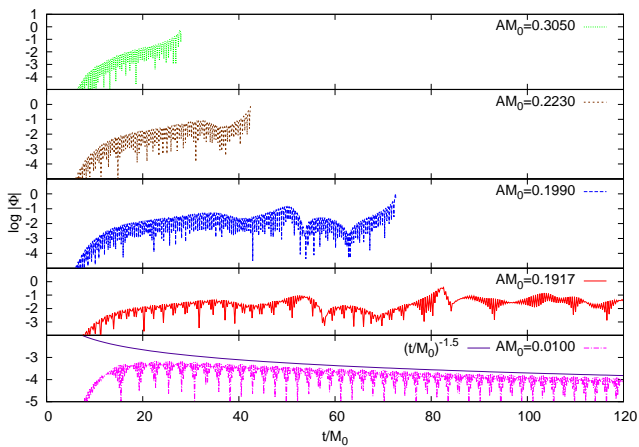,width=8.5cm}
\caption[]{Phases of a massive scalar collapse. The scalar field at the center of coordinates is shown as a function of time for selected decreasing amplitudes (from top to bottom). The first three upper panels correspond to the delayed collapse region of the diagram~\ref{fig:diagram}. The curves are truncated at the instant of an apparent horizon formation, which can occur after hundreds of reflections. The red curve ($AM_0=0.1917$) corresponds to a stable soliton star, cf. Fig.~\ref{fig:solitonstar}. As the amplitude is decreased further, the field decays with a characteristic $\sim t^{-3/2}$ fall-off, as shown by the magenta curve in the bottom panel. Initial data is specified by $r_0=5, w=1$ and $\mu=8$.
}
\label{fig:reflections}
\end{figure}

\begin{figure}[ht]
\psfig{file=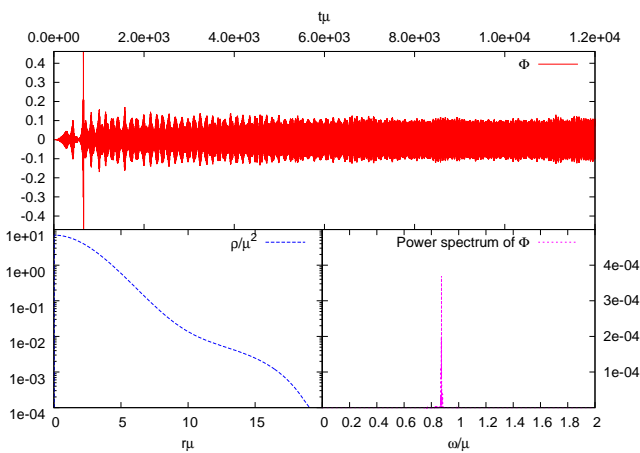,width=8.5cm}
\caption[]{Soliton star. Top panel: same as red curve ($AM_0=0.1917$) of Fig.~\ref{fig:reflections}, but now the evolution lasts for thousands of dynamical times. The configuration is stable up to at least $t\sim1.2\times10^4/\mu$. Bottom left panel: energy density of the compact configuration in the stationary regime. Bottom right panel: spectral decomposition showing one single dominant Fourier frequency.
}
\label{fig:solitonstar}
\end{figure}

Figure~\ref{fig:diagram} shows that the parameter space where $A<A_{*}^{\rm I}$ has much more structure. The possible outcomes of the time evolution are summarized in Figs.~\ref{fig:reflections} and~\ref{fig:solitonstar}, which show the scalar field $\Phi(0,t)$ for different initial amplitudes $A$. We find that for $A<A_{*}^{\rm I}$, the collapse can still occur, but is delayed by multiple reflections at the massive barrier. This situation is akin to the AdS case~\cite{Bizon:2011gg}, and we find it possible to tune the initial amplitude such that the number of reflections before collapse, and the collapse time, grow extremely large (possibly without bound).
Thus, in the type-I case, collapse to a black hole can occur promptly or after several successive reflections. For example, the middle panel (blue curve) in Fig.~\ref{fig:reflections} corresponds to hundreds of reflections before the field eventually collapses at $t\sim 75 M_0$.


It is tempting to conjecture that a confining mechanism similar to the AdS boundary is here in place: 
the mass term can trap arbitrarily small perturbations within a distance of the order of the Compton wavelength, and nonlinear interactions eventually lead to collapse, albeit possibly in ergodic time. 
Our results show otherwise, namely that the gravitational collapse halts for sufficiently small amplitudes. Indeed, we performed high-resolution ($dr/M_0=1/4000$) simulations lasting for up to tens of thousands of dynamical times, tracking the collapse time as a function of the amplitude. For a given mass $\mu$, we find another threshold value of the amplitude, 
\begin{equation}
 A_*^{\rm delayed}\sim \mu (w\mu)^{-1.2}\,, \label{eq:threshold_delayed}
\end{equation}
below which no collapse occurs. As $A\to A_*^{\rm delayed}$ the collapse time, as well as the number of bounces, becomes infinite and, when $A<A_*^{\rm delayed}$, the system approaches a stable bound-state configuration, as shown by the red curve in Fig.~\ref{fig:reflections} and detailed in Fig.~\ref{fig:solitonstar}. A detailed analysis of these solutions, including spectral content and spatial distribution of the energy density, suggests they belong to the family of oscillating soliton stars, or ``oscillatons'' described in Refs.~\cite{Seidel:1991zh,Seidel:1993zk,Page:2003rd}.
Our results confirm and extend the stability analysis performed in Ref.~\cite{Seidel:1991zh}, providing further evidence for the nonlinear stability of such solutions under gravitational collapse. Note, however, that oscillating soliton stars are only meta-stable, because they slowly decay through energy emission to infinity~\cite{Grandclement:2011wz}.

We thus find an interesting parallel with the AdS results: the mass term confines and favors collapse, but the existence of stable bound states where the field can relax into may prevent collapse to black holes.
This is highly reminiscent of the recently discovered stability islands in AdS~\cite{Buchel:2013uba,Maliborski:2013ula}.
In contrast with our case, the stability islands in AdS seem to form a compact set in the phase diagram, presumably because for very large spatial extent, the initial data ``bumps'' into the AdS boundary making the connection less clear.
The phase transition between (delayed) collapse and oscillatons explains the mass discontinuity reported in Ref.~\cite{Brady:1997fj} for type-I collapse. At variance with the type-II case, the collapse is not directly interfaced to the decay of the scalar field, and the threshold $A=A_*^{\rm delay}$ corresponds to the unstable branch of the mass-versus-radius curve of the soliton star~\cite{Brady:1997fj,Seidel:1991zh,Seidel:1993zk,Page:2003rd}. The mass discontinuity is indeed associated to the mass of the unstable oscillaton star.

As the amplitude decreases further, the characteristic wavelength of the soliton-star solutions increases. An important question is whether such solutions exist for arbitrarily small amplitude in some region of the parameter space. Our results indicate that, for sufficiently small amplitude, smaller than
\begin{equation}
 A_*^{\rm soliton}\sim \mu (w\mu)^{-2}\,,   \label{eq:threshold_soliton}
\end{equation}
the field generically dies away. This is shown by the magenta curve in the bottom panel of Fig.~\ref{fig:reflections}, which displays a clear $t^{-3/2}$ power-law decay, characteristic of massive fields for which Huygens principle is not valid~\cite{feshbach,Witek:2012tr}. 

A precise characterization of the transition region between type-I and type-II collapse is more challenging and would depend on the initial data. However, a solid qualitative outcome of our simulations is that soliton stars only exist above a certain value of the dimensionless coupling, $w\mu\gtrsim {\cal O}(1)$. This implies the existence of a \emph{triple point} in the phase diagram, which separates black-hole formation, soliton stars and power-law decay. Another interesting result is the fact that the area of the parameter space for soliton stars increases in the $w\mu\gg1$ limit. In the rightmost part of the diagram~\eqref{fig:diagram} even a very small initial amplitude would form a stable bound state rather than dispersing at infinity.

\section{Discussion and Outlook}
The discovery of the nonlinear instability of AdS taught us how important confinement is in a gravity theory. Our analysis confirms that a confining mechanism alone is not sufficient to produce gravitational collapse with generic initial data; the absence of stable, nonlinear solutions other than black holes is another crucial ingredient. In the AdS case, such solutions are the exception rather than the rule~\cite{Buchel:2013uba,Maliborski:2013ula}. In the case of self-interacting scalar fields in asymptotically flat spacetime, the existence of soliton stars generically provides a competitive alternative to collapse. 

Our results also show how important the role of dissipation is. An obvious difference between the confining mechanisms provided by the AdS asymptotics and by a mass term is that the latter can only confine low-frequency waves at linear level~\cite{Cardoso:2005vk,Dolan:2007mj,Dolan:2012yt,Pani:2012vp,Pani:2012bp,Witek:2012tr}. Thus, if the collapse mechanism were due to turbulent blueshift of energy alone, it is conceivable that nonlinear effects would be quenched after some point and that energy would slowly leak away. In other words, the existence of a finite-height potential barrier imposes limits on the effectiveness of turbulent effects as a source of gravitational collapse. Nevertheless, we see no hints of blueshift. On the contrary, the frequency of the dynamical quantities is roughly constant or even decreasing in time. Dissipation at the boundary, through the leakage of energy to infinity, seems to be essential to quench nonlinearities and to halt the collapse in the asymptotically flat case. 

Nonspherical initial data will also dissipate through gravitational-wave emission at linear level on short timescales.
But this mechanism presumably leaves behind a spherical configuration which can then evolve nonlinearly according to the phase diagram of Fig.~\ref{fig:diagram}. In this sense, our study describes the late-time behavior of generic nonspherical configurations, after linear gravitational-wave damping has taken its toll.

In order to study the collapse of arbitrarily small initial data, accurate simulations have to be evolved for long time.
Although our simulations suggest that a region of dispersion to infinity is always present in the phase diagram~\ref{fig:diagram}, further analytical insights would be crucial to provide a definitive answer to the nonlinear stability problem of soliton stars. Likewise, we cannot exclude that after a period of power-law decay, the system approaches a stable configuration or that eventually it collapses on ergodic time scales.

One might speculate on the astrophysical relevance of stable soliton stars~\cite{Seidel:1991zh}. The threshold~\eqref{eq:threshold_soliton} translates into $M_0 \mu \gtrsim 0.02 (w\mu)^{-1}$. For an hypothetical stable massive field with mass $\sim{\rm eV}$, the critical mass for soliton-star formation is of the order of $10^{-12}(w\mu)^{-1} M_\odot$, and only very extended initial configurations would provide interesting astrophysical scenarios. 

We studied in detail the case of massive scalar fields, but a less extensive search for $\Phi^4$ interactions yields qualitatively similar results. Soliton-star solutions exist also in the case of massive scalar fields with $\Phi^4$ self-interactions~\cite{Honda:2001xg,ValdezAlvarado:2011dd}. A more detailed analysis would be relevant, for instance, to understand the role of gravitational collapse in some inflationary models, where the massive inflaton is described by a $\Phi^4$ self-interaction.
Our results suggest that a very rich phenomenology might exist also in other confining geometries like, for example, in compact stars. In this case fluid perturbations are confined within the stellar radius and the secular stability of these objects is an interesting open problem.

\begin{acknowledgments}
 We appreciate useful discussions with Juan Degollado, \'Oscar Dias, Roberto Emparan, Gyula Fodor, Akihiro Ishibashi, 
Luis Lehner, Maciej Maliborski, Jos\'e Nat\'ario, Andrzej Rostworowski, Jorge Santos and all participants of the
``NR/HEP2 Spring School'' in Lisbon, of the YITP-T-11-08 workshop on ``Recent
advances in numerical and analytical methods for black hole dynamics,''
the Perimeter Institute workshop ``Exploring AdS-CFT Dualities in Dynamical Settings,'' and the ``MM13'' meeting in Mons.
We acknowledge financial support provided under the European Union's FP7 ERC Starting Grant ``The dynamics of black holes:
testing the limits of Einstein's theory'' grant agreement no. DyBHo--256667 and through the Intra-European Marie Curie contract aStronGR-2011-298297.
This research was supported in part by Perimeter Institute for Theoretical Physics. 
Research at Perimeter Institute is supported by the Government of Canada through 
Industry Canada and by the Province of Ontario through the Ministry of Economic Development 
$\&$ Innovation.
This work was also supported by the NRHEP 295189 FP7-PEOPLE-2011-IRSES Grant, and by FCT-Portugal through projects
CERN/FP/123593/2011 and IF/00293/2013.
Computations were performed on the ``Baltasar Sete-Sois'' cluster at IST and
on the ``venus'' cluster at YITP.
\end{acknowledgments}


\end{document}